# Improving the resolution of Cryo-EM single particle analysis

Zhenwei Luo

Rice University, Department of Bioengineering
Houston, TX
`Zl24@rice.edu`

**Abstract.** We presented a new 3D refinement method for Cryo-EM single particle analysis which can improve the resolution of final electron density map in this paper. We proposed to enforce both sparsity and smoothness to improve the regularity of electron density map in the refinement process. To achieve this goal, we designed a novel type of real space penalty function and incorporated it into the refinement process. We bridged the backprojection step with local kernel regression, thus enabling us to embed the 3D model in reproducing kernel Hilbert space using specific kernels. We also proposed a first order method to solve the resulting optimization problem and implemented it efficiently with CUDA. We compared the performance of our new method with respect to the traditional method on real datasets using a set of widely used metrics for Cryo-EM model validation. We demonstrated that our method outperforms the traditional method in terms of those metrics. The implementation of our method can be found at https://github.com/alncat/cryoem.

## 1    Introduction

Cryo-electron microscopy (Cryo-EM) single particle analysis becomes an increasingly popular method to visualize molecular structure. Cryo-EM has certain advantages over the traditional X-ray crystallography as it doesn't require crystallization and isn't plagued by the phasing problem. However, there are also many new challenges arise in this promising technique. The central problem of Cryo-EM single particle analysis is the incompleteness of experimental observations. More specifically, the relative orientations and translations of all particles are missing, and the memberships of all particles in a structural heterogenous dataset remain unknown. Moreover, the signal to noise ratio of Cryo-EM dataset is often very low since the electron exposure of the sample needs to be strictly limited to prevent radiation damage [1], [2]. Other problems often present in Cryo-EM dataset is the nonuniform angular sampling, which results in inadequate samples or even no samples in certain orientations [3]. Therefore, building an ultrahigh dimension 3D model with incomplete and highly noisy data is ill-posed. To alleviate this problem, prior assumptions must be incorporated into the building process to ensure the uniqueness of solution and the objectivity of final model.

Two prominent features of a 3D molecular model are sparse and smooth. More specifically, since the electron densities of a molecule only occupy a small part of the 3D volume it resides in, the molecule models are often sparse; Because atoms in molecules are connected through chemical bonds, the electron densities of molecules vary smoothly across the space. The importance of smoothness prior is widely recognized in Cryo-EM based 3D model refinement. Early attempt to enforce the smoothness of the density map is by applying Wiener filter [4]. Later approaches improve upon the Wiener filter by using Bayesian statistics. As it was proposed by Scheres *et al.*, he first assumed that the Fourier components of the density map are distributed according to a gaussian distribution [5]. He then obtained a closed form solution of the corresponding Maximum a posterior (MAP) problem for the Fourier components of the density map, which is similar to the wiener filtering, and employed this formula to update the 3D model in maximization step. This approach is referred to as the traditional approach in the rest of this paper. Though sparsity is a popular prior in solving inverse problem, it is a relatively novel notion to the Cryo-EM based 3D model refinement. In this paper, we propose a new approach to regularize the 3D model which encourages both sparsity and smoothness of the 3D model. Our approach is inspired by many successful methods which are proposed recently for solving the ill-posed inverse problem by imposing the sparse and smooth priors, such as $l_1$ regularization, compressed sensing and total variation [6]–[9]. It is tempting to apply the state-of-the-art sparsity learning algorithms to the Cryo-EM based 3D model reconstruction problem. In this paper, to encourage the sparseness and smoothness of the reconstructed electron density map while suppressing bias, we proposed a non-concave non-smooth restraint by combing lasso and total variation. Since the non-concave non-smooth target function is difficult to optimize directly, we designed a reweighting scheme to approximately optimize the target function with a sequence of weighted $l_1$ regularization and total variation problem. The weighted problem can be solved by the smoothing proximal gradient method. As it is later revealed in the method section, the traditional approach can be viewed as applying a translation invariant rotationally symmetric kernel to the 3D model, whereas our new approach applies a spatially varying kernel to the 3D model. Real 3D models often present anisotropic smoothness, i.e., electron densities remain constant along the directions of chemical bonds while they decay fast in directions perpendicular to the directions of chemical bonds. Therefore, our new approach can adapt to spatially varying smoothness exists in the 3D models of macromolecules and outperform traditional approach in certain classes of models. Another challenge of 3D model reconstruction in real space is its ultrahigh dimensionality which results in prohibiting computational cost. For example, a common $512 \times 512 \times 512$ 3D volume represents hundreds of millions of variables. We address this challenge by implementing a CUDA accelerated solver for our method. It has been shown that our solver can obtain medium accurate solution within 10 seconds on a problem of size $512 \times 512 \times 512$. Our final modification to the Cryo-EM refinement process is bridging the backprojection with local kernel regression, thus paving a novel way to promote the smoothness of 3D model. We propose using gaussian kernel in the backprojection step which can represent a widely used reproducing kernel Hilbert space—a Hilbert space endowed with some notion of smoothness.

In the following paper, we first describe the underlying theory of Cryo-EM structure determination to define the statistical framework on which we later work. We then propose a new type of restraint for the density map in real space which encodes the available prior information. Next, since there is no closed form solution for the new target function, we design an iterative method to compute a 3D model maximizes the new target function. Using tools in real analysis, we elucidate the difference between the traditional method and our new method theoretically. We also derive the relationship between backprojection and local kernel regression and propose a new choice of kernel. Finally, we demonstrate the effectiveness of our new method by applying it to real datasets and compare the refined results with those obtained by traditional approach. We can observe improvements in both the gold standard FSC and model-map FSC in these datasets, which suggests that our method is able to obtain reconstructed maps with higher resolution.

## 2    Methods

The Cryo-EM refinement problem can be formulated as fitting a generative model whose data generation process is illustrated as follows. For simplicity, we assume the particles are from a single structure, while structural heterogeneity can be easily incorporated into our generative model by treating the class membership of each particle as a hidden variable [5]. The images collected in Cryo-EM experiments are 2D projections of a 3D molecular structure. The fourier transform of the image has following relation with the fourier transform of the 3D molecular structure. Let the fourier transform of a 3D molecular structure be $V$, we first arrange the 3D volume $V$ into a vector with $L$ elements. Assume an $N \times N$ image $i$ is formed by projecting the 3D volume $V$ with the euler angle set $\phi$ and shifting by $[\Delta_x, \Delta_y]$ from the origin, using projection-slice theorem, the fourier transform of the image $i$ can be expressed as

$$X_{i,j} = e^{-i\frac{2\pi}{N}(\Delta_x k + \Delta_y l)} \text{CTF}_{ij} \sum_{l=1}^{L} P_{jl}^{\phi} V_l, \tag{1}$$

where $X_{ij}$ is the $j$th component of the fourier transform of the image $i$ whose corresponding 2D index is $[k, l]$, $\text{CTF}_{ij}$ is the $j$th component of the contrast transfer function for the image $i$ [10], [11], and $P_{jl}^{\phi}$ is the slice operator which takes out the plane in the 3D fourier transform $V$ which is rotated from the $xy$ plane according to the euler angle set $\phi$. Note that images acquired in experiments are often contaminated by noises, we can account the uncertainty of experimental data by a distribution. Suppose the fourier component $X_{ij}$ is distributed according to gaussian with the mean defined in equation (1) and variance $\sigma^2$, and the gaussian noise of each component is independent, the marginal probability of observing image $i$ can be obtained by integrating out all possible orientations $\phi$ and translations $\Delta = [\Delta_x, \Delta_y]$ as following,

$$P(X_i|V) \propto \int_{\phi, \Delta} \exp\left\{-\frac{1}{\sigma^2} \sum_{j=1}^{J} \left(X_{ij}\right.\right.$$
$$\left.\left. - e^{-i\frac{2\pi}{N}(\Delta_x k + \Delta_y l)} \text{CTF}_{ij} \sum_{l=1}^{L} P_{jl}^{\phi} V_l\right)^2\right\} d\phi d\Delta. \tag{2}$$

In following sections, we omit translation factors in the squared difference term in equation (2) to simplify expressions. After obtaining the marginal probability of data, we can then reconstruct the 3D model by maximizing the total log likelihood function

$$\sum_{i=1}^{N} \log P(X_i|V). \tag{3}$$

However, this log likelihood function fits the model to experimental data without applying any prior, thus often generating overfitted models in practice. Hence, we should consider adding appropriate priors to the log likelihood function to guarantee the feasibility of solution and reduce overfitting during refinement. As mentioned in introduction, the 3D molecular models are both sparse and smooth. In order to incorporate these priors into refinement, a mathematical formulation for these priors must be developed. On the one hand, mathematician has associated the smoothness of a function with the norm of its gradient. On the other hand, sparsity refers to the number of zeros in the values of function [12]. In the following section, we will formulate different smoothness priors and reveal their difference. Two key equations illustrating the effect of previous smoothness restraint and our new smoothness restraint are equation (7) and equation (19), respectively.

Only considering smoothness so far, let $x$ be the 3D model, we can promote the smoothness of a 3D model by maximizing the log likelihood while restraining the gradients of solution, namely, maximizing the following target function

$$\sum_{i=1}^{N} \log P(X_i|V) - \lambda \|\nabla x\|_p^p, \tag{4}$$

where $\lambda$ and $p$ are positive parameters. We start by demonstrating that the restraint on fourier coefficients in traditional method closely resembles the squared norm of gradients. When $p = 2$, the norm of gradients can be simply transformed to a weighted norm of fourier coefficients using certain results in harmonic analysis. Since the fourier transform of the derivative of a function equals with multiplying the fourier transform of this function by the corresponding frequency component, that is, $\mathcal{F}\left(\frac{\partial x}{\partial i}\right) = h\hat{x}$, using Plancherel's theorem, we have

$$\|\nabla x\|_2^2 = 4\pi^2 \sum_{hkl} (h^2 + k^2 + l^2) |\hat{x}_{hkl}|^2, \tag{5}$$

where $\hat{x}_{hkl}$ represents the fourier transform of the 3D model $x$. Traditional method proposed in [13] enforces smoothness by applying a quadratic restraint on the magnitudes of fourier transforms based on the assumption that they are distributed according to gaussian. In traditional method, the regularization parameter $\lambda$ depends only on the radius of fourier shell, i.e., all fourier transforms $\hat{x}_{hkl}$ with the same radius $\sqrt{h^2 + k^2 + l^2} = r$ are regularized by the same regularization parameter. More specifically, the regularization strength $\lambda(r)$ is the product of the average of the factor $N$ appeared in equation (28) in that shell and a function $w(r)$ of the inverse of the gold standard FSC value of that shell [14], that is, $w(r) = \frac{1}{\text{FSC}(r)} - 1$, where $\text{FSC}(r)$ is the value of FSC of the corresponding shell. In fact, $w(r) = \frac{1}{\text{SNR}(r)}$, that is, the weight is the inverse of the signal to noise ratio at that shell according to [15], [16]. It becomes evident that traditional method enforces smoothness in a similar way as restraining the squared norm of the gradient of 3D model $x$ since they are both functions of the radius of fourier shell. However, traditional method improves upon the simple squared gradient norm restraint by adopting the gold standard FSC derived restraint strength across different shells. To study the effect of the restraint of traditional method in more detail, we will examine how this restraint acts in the optimization process. By using gradient ascent algorithm, the penalized log likelihood function in equation (4) can be maximized by updating the fourier transform of 3D model $V$ as

$$V'_{hkl} = V_{hkl} + l\left(\sum_{i=1}^{N}\frac{\partial}{\partial V_{hkl}}\log P(X_i|V) - c(r)V_{hkl}\right). \qquad (6)$$

We can then understand the effect of the restraint by taking inverse fourier transform for the term $c(r)V_{hkl}$. By convolution theorem, let the fourier transform of $c(r)$ be $K(\|u\|_2)$, which is a radial function, the inverse fourier transform of $c(r)V_{hkl}$ can be written as,

$$x'(u) = \int_{\mathbb{R}^3} K(\|u - v\|_2)x(v)dv. \qquad (7)$$

Equation (7) clearly states that the restraint in traditional method acts a translation invariant rotationally symmetric kernel $K(\|u - v\|_2)$ during the optimization process. At each step, the old solution is updated by a linear combination of the gradient of log marginal likelihood function and the radial kernel smoothed old solution. We thus postulate that the traditional method biases toward the 3D model with homogeneous smoothness across space.

We will present our new priors in the remainder of this section. Assume $x$ is the 3D volume which is rearranged into a vector, that is, a grid point with index $[i, j, k]$ is mapped to the $h$th component $x_h$ of $x$, and let $A$ be the corresponding 3D Fourier transform matrix, we can express the Fourier coefficients of the 3D volume $V$ as the result of matrix vector multiplication, namely, $V = Ax$. By encoding the smoothness of model using total variation norm, i.e., setting $p = 1$ in equation (4), and the sparsity of the model using $l_1$ norm, the log likelihood with priors is of the form

$$\sum_{i=1}^{N}\log P(X_i|V) - \alpha\|x\|_1 - \beta\|\nabla x\|_1, \qquad (8)$$

where $\|x\|_1$ is the $l_1$ norm, $\|\nabla x\|_1$ is the total variation norm, which is the sum of the magnitude of gradient at each grid point, $\|\nabla x\|_1 = \sum_{i=1}^{L}\|\nabla x_i\|_2$ and $\alpha$ and $\beta$ are positive parameters. The gradient $\nabla x_i$ in the total variation norm can be calculated via discrete approximation. Though these two priors can effectively guarantee both sparsity and smoothness of the solution, they will introduce certain biases to the final solution as described in [17]. For example, the solution obtained with $l_1$ regularization tends to shrink the nonzero elements. Fan discovered that nonconcave penalty can effectively prevent the true nonzero elements from overly shrinking while preserving sparsity [17]. Hence, we consider employing a nonconcave penalty—the log norm which is advocated in [18] to reduce possible biases in solution. Our log likelihood function with nonconcave priors can be written as

$$\sum_{i=1}^{N} \log P(X_i|V) - \sum_{j=1}^{L} (\alpha \log(|x_j| + \epsilon) + \beta \log(\|\nabla x_j\|_2 + \epsilon)) \tag{9}$$

where $\epsilon$ is a small positive constant introduced to guard against the singularity of log function around zero.

The remainder of this section is devoted to proposing an algorithm to optimize our likelihood function with newly proposed priors. The total penalized likelihood function consists of three different terms which can be tackled by different algorithms. First of all, the log likelihood function of the marginal probability in equation (9) can be optimized by the expectation-maximization method [19]. This algorithm works as follows; Since the difference between log likelihoods of the marginal probability can be lower bounded by the difference between the sums of log likelihoods of the joint probability weighted by their corresponding posterior probabilities for latent variables, i.e., $\log P(X_i|V) - \log P(X_i|V_0) \geq \sum_\phi -P(\phi|X_i, V_0)(\|X_i - \mathrm{CTF}_i P^\phi V\|^2 - \|X_i - \mathrm{CTF}_i P^\phi V_0\|^2)$, maximizing the lower bound improves the log likelihood of the marginal probability at least as much [20]. At the expectation step, we calculate the posterior probability of latent variables conditioned on a given image and a model. The method to compute the posterior probability derived in [5] can be applied in the context of our method without any modification. At the $k$th maximization step, we maximize the log likelihoods of marginal probabilities by replacing them with the lower bounds, as follows:

$$\max_x \sum_{i=1}^{N} \sum_\phi -P(\phi|X_i, V_{k-1}) \|X_i - \mathrm{CTF}_i P^\phi V\|^2 \tag{10}$$
$$- \sum_{j=1}^{L} (\alpha \log(|x_j| + \epsilon) + \beta \log(\|\nabla x_j\|_2 + \epsilon)),$$

where $P(\phi|X_i, V_{k-1})$ is the conditional probability of the latent variables given the observation $X_i$ and the model $V_{k-1}$ from the previous iteration, $\alpha$ and $\beta$ are the weights of $l_1$ norm and total variation norm, respectively. The 3D model at the $k$th step thus is the maximizer of equation (10).

The nonsmoothness and nonconcavity of newly introduced penalties pose big challenges for optimizing the corresponding objective function. To address the nonconcavity of log function, we use a sequence of weighted concave loss to approximate the nonconcave loss as in [18], [21], [22]. More specifically, at iteration $i + 1$, let the solution obtained in previous iteration be $x^i$, we approximate equation (10) with the weighted $l_1$ and total variation norms and define the solution at this iteration as the maximizer of the following equation,

$$\max_x \sum_{l=1}^N \sum_\phi -P(\phi|X_l, V_{k-1}) \left\| X_l - \text{CTF}_l P^\phi V \right\|^2 \tag{11}$$
$$- \sum_{j=1}^L (\alpha \frac{|x_j|}{|x_j^i| + \epsilon} + \beta \frac{\left\| \nabla x_j \right\|_2}{\left\| \nabla x_j^i \right\|_2 + \epsilon}).$$

These approximations are the tangent lines for each logarithm in the log norm at the previous solution $x^i$. Since the log function is a concave function, its tangent line are its upper bounds [18]. Minimizing the upper bounds surely leads to lower log norm since the log function is monotone. Therefore, this approximation algorithm is guaranteed to maximize the penalized likelihood. It's easy to see that the weight norms have similar debiasing effects as the log function since it downweighs the components with large values in the $l_1$ or TV norms, and it is of the same spirit as the adaptive lasso method proposed by Zou [21]. By decomposing the nonconcave problem as a series of concave optimization problems, we may leverage existing algorithms such as gradient ascent in concave optimization to solve the concave subproblem. However, our target function contains some nonsmooth terms, namely, the total variation (TV) norm and the $l_1$ norm in equation (11). These terms are nondifferentiable around zero and prevent us to apply the gradient ascent algorithm directly to maximize the penalized likelihood function. We then leverage algorithms for optimizing nonsmooth function to deal with those terms. For the total variation norm, we propose to use Nesterov smoothing to obtain an approximate gradient [23], which allows us to circumvent the nondifferentiability of TV norm and apply gradient ascent. To derive the gradient of Nesterov smoothed TV norm, we begin with giving the discrete form of TV norm. For a point $x[i, j, k]$ of a 3D model $x$, the discrete approximation of the gradient of $x$ at this point is of the form

$$Dx[i, j, k] = \begin{bmatrix} D_1 x[i, j, k] \\ D_2 x[i, j, k] \\ D_3 x[i, j, k] \end{bmatrix} = \begin{bmatrix} x[i, j, k] - x[i-1, j, k] \\ x[i, j, k] - x[i, j-1, k] \\ x[i, j, k] - x[i, j, k-1] \end{bmatrix}. \tag{12}$$

Denote $D_i$ as the matrix form of the discrete differentiation operator along the $i$th dimension for the vectorized volume $x$, namely, we have $D_i x$ as the gradient of $x$ along the $i$th dimension, and let $D = [D_1, D_2, D_3]^*$ be a matrix composed by concatenating $D_i$ by rows, the dual norm of TV norm can be defined as,

$$\|x\|_{\text{TV}} = \max_{u \in Q_d} \langle u, Dx \rangle, \tag{13}$$

where $u = [u_1, u_2, u_3]^* \in Q_d$ is a vector formed by concatenating three vectors of the same size as $x$ and $Q_d$ is defined as the vector space of the same dimension as $u$ which satisfies the inequality $u_1[i,j,k]^2 + u_2[i,j,k]^2 + u_3[i,j,k]^2 \leq 1$. Using Nesterov smoothing with smoothing parameter $\mu$, we can obtain a new functional $f_\mu(x)$, i.e., the smoothed total variation norm, which is of the form,

$$f_\mu(x) = \max_{u \in Q_d} \langle u, Dx \rangle - \frac{\mu}{2} \|u\|_2^2, \tag{14}$$

[24]. The gradient of the smoothed TV norm $f_\mu(x)$ can be written as

$$\nabla f_\mu(x) = D^* u_\mu(x), \tag{15}$$

where $u_\mu(x)$ is a vector of the form $[u_1, u_2, u_3]^*$ and for each dimension $a \in [1,2,3]$,

$$u_a[i,j,k] = \begin{cases} \mu^{-1}(D_a x)[i,j,k], \text{if } \|\nabla x[i,j,k]\| < \mu \\ \|\nabla x[i,j,k]\|^{-1}(D_a x)[i,j,k], \text{otherwise.} \end{cases} \tag{16}$$

The above derivation simply shows that the gradient of smoothed TV norm can be obtained by first calculating the norm of discrete gradient of the volume $x$ at each point $[i,j,k]$, and then setting the gradient norm of the volume $x$, $\|\nabla x[i,j,k]\|$, with value smaller than the smoothing parameter $\mu$ to the smoothing parameter itself, thus keeping the denominator of the gradient of TV norm in a valid range and avoiding the nondifferentiability of the nonsmoothed TV norm around zero. We pause to expose the effect of TV norm in the optimization process by examining the gradient $\nabla f_\mu(x)$. For a grid point $(i,j,k)$, the gradient at this point is of the form,

$$\nabla f_\mu(x)_{ijk} = \sum_{a=1}^{3} u_a[i,j,k] - u_a[(i,j,k) + \Delta_a], \tag{17}$$

where $\Delta_a$ is a $1 \times 3$ vector with one on the $a$th entry and zeros elsewhere. Substituting equation (16) into equation (17), we can observe that the gradient of TV norm at grid point $(i, j, k)$ depends on gradients of 3D model around this point. More specifically, equation (17) can be reformulated as

$$\nabla f_\mu(x)_{ijk} = c_{(i,j,k)} x[i, j, k] - \sum_{a=1}^{3} c'_{(i,j,k) \pm \Delta_a} x[(i, j, k) \pm \Delta_a], \tag{18}$$

where $c_{(i,j,k)} = 3 \|\nabla x[i,j,k]\|^{-1} + \sum_{a=1}^{3} \|\nabla x[(i,j,k) + \Delta_a]\|^{-1}$, $c'_{(i,j,k) - \Delta_a} = \|\nabla x[i,j,k]\|^{-1}$ and $c'_{(i,j,k) + \Delta_a} = \|\nabla x[(i,j,k) + \Delta_a]\|^{-1}$. The continuous form of equation (18) can be written as

$$x'(u) = \int_{\mathbb{R}^3} K(u, v) x(v) dv, \tag{19}$$

where the kernel $K(u, v)$ depends on the location $u$ in smoothed volume and the location $v$ in original volume simultaneously. Therefore, the TV norm acts as a spatially varying kernel $K(u, v)$ during the optimization process, which can adapt to heterogenous smoothness exists in the 3D model.

We continue to design an optimization algorithm for the $l_1$ regularization part. The target function with $l_1$ norm penalty can be optimized by proximal operator, which is the soft-thresholding operator in this case [25]. Combining above techniques, denote $l$ as the expected minus log likelihood $\sum_{i=1}^{N} P(\phi | X_i, V_0) \|X_i - \text{CTF}_i P^\phi V\|^2$, $w_j$ as the weight $\frac{1}{\|\nabla x_j^i\|_2 + \epsilon}$ in TV norm, $w_j'$ as the weight $\frac{1}{|x_j^i| + \epsilon}$ in $l_1$ norm, and $x^i$ as the 3d volume at iteration $i$ of this maximization step, and let the learning rate be $l_r$, at $i + 1$th iteration, for the $j$th component of $x$, to solve the optimization problem defined in equation (11), the proximal operator to update the 3d volume is defined as following,

$$x_j^{i+1} = \min_{x_j} \frac{1}{2} \left\| x_j - (x_j^i - l_r \left( \nabla l(x_j^i) + \beta w_j \nabla f_\mu(x_j^i) \right)) \right\|_2^2 \tag{20}$$
$$+ l_r \alpha w_j' \|x_j\|_1,$$

which has a closed form solution as follows,

$$x_j^{i+1} = \begin{cases} 0, |x_j^{i'}| < l_r \alpha w_j' \\ x_j^{i'} - l_r \alpha w_j' \text{sign}(x_j^{i'}), |x_j^{i'}| \geq l_r \alpha w_j' \end{cases} \tag{21}$$

where $x^{i'} = x^i - l_r \left( \nabla l(x^i) + \beta w_j \nabla f_\mu(x^i) \right)$ is the 3D model after gradient descent update and sign represents the sign function. We can clearly see the effect of $l_1$ norm here as it sets the volume where the value is relatively small to zero and guarantees the sparsity of the volume. Since the electron densities for molecules should have higher values comparing with the background noises, we expect the background noises will be suppressed while the electron densities of molecules will keep untouched after applying the weighted $l_1$ restraint. The sparsity restraint results in cleaner 3D models.

We also consider using the implicit gradient descent method proposed in [26] to improve the stability of optimization process. This can be achieved by adding a quadratic restraint which favors a new solution that is closer to the solution obtained in previous maximization step. Therefore, at the $i + 1$th iteration of the $k$th maximization step, the optimal model $x$ is defined as the maximizer of the equation below,

$$\max_x \sum_{l=1}^{N} -P(\phi|X_l, V_{k-1}) \left\| X_l - \mathrm{CTF}_l P^{\phi} V \right\|^2 \tag{22}$$
$$- \alpha \sum_{j=1}^{L} \frac{|x_j|}{|x_j^i| + \epsilon} - \beta \sum_{j=1}^{L} \frac{\left\| \nabla x_j \right\|_2}{\left\| \nabla x_j^i \right\|_2 + \epsilon} - \gamma \|x - x^{k-1}\|_2^2,$$

where $x^{k-1}$ is the solution obtained at the $k - 1$th maximization step and $\gamma > 0$ is the weight of the implicit gradient descent restraint. Since the newly added term is quadratic and differentiable, it contributes a new term to the gradient descent updated model $x^{i'}$. In conclusion, at each maximization step, we use equation (21) iteratively to update the 3d volume $x$ and obtain a final solution after certain number of iterations.

In our new method, we apply the expectation-maximization method with the aforementioned modifications to reconstruct the 3d volume until the convergence criteria are met.

## 2.2 Local Kernel Regression

An important step in the Cryo-EM refinement is backprojection. After the orientation and translation of an image is estimated, the 2D fourier transform of image can be put back to its corresponding plane in the 3D fourier transform of electron density map in this step. Backprojection is used to construct the lower bound of the log marginal likelihood of each image as stated before, and can be formulated as an optimization problem where the 3D volume obtained by minimizing the total loss between slices of 3D volume and the corresponding fourier transformations of 2D images, i.e.,

$$\min_V \sum_{i=1}^{N} \sum_{\phi} P(\phi|X_i, V_0) \left\| X_i - \mathrm{CTF}_i P^{\phi} V \right\|^2, \tag{23}$$

which is the first term in equation (10). A practical problem to be considered here is that the 3D volume is discretely sampled with integer indices, while the 2D slice corresponding to the backprojected 2D image has noninteger indices. Formally, let $[i, j] \in \mathbb{Z}^2$ be the index of a point in the 2D image, suppose the corresponding slice of the 2D image in 3D volume is defined by the latent parameter $\phi$, which is a set of euler angles defining the rotation of plane if ignoring translation, after backprojection, the index of the point in 3D volume $[h, k, l]$ is of the form,

$$\begin{pmatrix} h \\ k \\ l \end{pmatrix} = R_\phi \begin{pmatrix} i \\ j \\ 0 \end{pmatrix},$$

(24)

where $R_\phi$ is the rotation matrix transforming the 2D image according to the euler angles $\phi$. Obviously, $[h, k, l] \in \mathbb{R}^3$, while the sampling point of the 3D volume has index $[h, k, l] \in \mathbb{Z}^3$. Therefore, there is no one to one mapping between the sampling point in the backprojected 2D slice $P^\phi V$ and the sampling point in the 3D volume $V$. This discrepancy is solved by using the nonparametric regression, which has been proposed in image processing by [27]. Simply put, denote $Y_i$ as the value at a certain point $x_i \in \mathbb{R}^N$, and let y be the value at the point $x \in \mathbb{R}^N$ which is to be predicted, as in [27], [28], we can define y as the minimizer of

$$\sum_{i=1}^{L} K \left( \frac{\|x - x_i\|}{h} \right) \|Y_i - y\|^2 ,$$

(25)

where $K(\cdot)$ represents a chosen kernel function and $h$ is the bandwidth of the kernel function. Returning to the context of cryo em refinement, for an orientation $\phi$, let the corresponding grid point of a fourier coefficient $j$ of an image $X_i$ in a 3D volume $V$ be $n_j(\phi) = [h_j, k_j, l_j]$ with $[h_j, k_j, l_j] \in \mathbb{R}^3$, the value of the grid point $n = [h, k, l] \in \mathbb{Z}^3$ is the minimizer of

$$\sum_{j=1}^{L} P(\phi|X_i, V_0) K \left( \frac{\|n_j(\phi) - n\|}{h} \right) \|X_{ij} - \mathrm{CTF}_{ij} V_{hkl}\|^2.$$

(26)

The total loss is of the form,

$$\sum_{hkl} \sum_{i=1}^{N} \sum_{j=1}^{L} \sum_{\phi} P(\phi|X_i, V_0) K \left( \frac{\|n_j(\phi) - n\|}{h} \right) \|X_{ij} - \mathrm{CTF}_{ij} V_{hkl}\|^2.$$

(27)

For our new method, we consider using the gaussian kernel as our kernel function, namely, $K\left(\frac{\|n_j-n\|}{h}\right) = \exp(-\frac{(h_j-h)^2+(k_j-k)^2+(l_j-l)^2}{h})$, and we set the bandwidth $h$ to be 2. For practical consideration, the fourier coefficient $X_{ij}$ should not be back projected to every grid point. Let $\llbracket \cdot \rrbracket$ be the round operator, we consider back projecting $X_{ij}$ with index $[h_j, k_j, l_j]$ to it neighboring grid points $[\llbracket h_j \rrbracket + \Delta_h, \llbracket k_j \rrbracket + \Delta_k, \llbracket l_j \rrbracket + \Delta_l]$, where $\Delta_a \in [-1,0,1]$. In the original implementation of *Relion*, they chose a kernel with trilinear interpolator like weight. Using the formalism introduced in this paper, their kernel function are of the form $K\left(\frac{\|n_j-n\|}{h}\right) =$ $w(h_j, h)w(k_j, k)w(l_j, l)$, where $w(x_j, x) = \begin{cases} x+1-x_j, x \leq x_j \\ x-x_j, x > x_j \end{cases}$, and $h \in$ $[\lfloor h_j \rfloor, \lceil h_j \rceil], k \in [\lfloor k_j \rfloor, \lceil k_j \rceil], l \in [\lfloor l_j \rfloor, \lceil l_j \rceil]$, where $\lfloor \cdot \rfloor$ and $\lceil \cdot \rceil$ are floor and ceiling operators, respectively. Hence the fourier coefficient $X_{ij}$ is backprojected to 8 neighboring grid points in the original implementation of *Relion*. We here take a second look at the equation (27). By grouping loss terms w.r.t $V_{hkl}$ into a single term and ignoring constant terms pertaining to $X_{ij}$, we can rearrange equation (27) as

$$N(n) \left\| V_{hkl} - \frac{\sum_{i=1}^{N}\sum_{j=1}^{L}\sum_{\phi}P(\phi|X_i,V_0)K\left(\frac{\|n_j(\phi)-n\|}{h}\right)\text{CTF}_{ij}X_{ij}}{N(n)} \right\|^2, \tag{28}$$

where $N(n) = \sum_{i=1}^{N}\sum_{j=1}^{L}\sum_{\phi}P(\phi|X_i,V_0)K\left(\frac{\|n_j(\phi)-n\|}{h}\right)\text{CTF}_{ij}^2$. Thus our new algorithm estimates the value of 3D volume at a grid point $[h,k,l]$ by a discrete approximation of the continuous convolution between kernel function and the weighted experimental data $V_{h'k'l'}$, which is of the form

$$V_{hkl} = \sigma \int_{h'k'l'} K\left(\frac{\|n'-n\|}{h}\right)V_{h'k'l'}dh'dk'dl', \tag{29}$$

where $\sigma$ is the normalization factor for the kernel. We can then choose a certain kernel to embed $V_{hkl}$ in the reproducing kernel Hilbert space (RKHS) associated with that kernel, which enforces $V_{hkl}$ to exhibit the smoothness of RKHS induced by the chosen kernel [29].

## 2.3    Gold Standard FSC

The gold standard Fourier Shell Correlation (FSC) is a method to determine the resolution of refined model without overfitting [14]. The FSC between the Fourier components $F$ and $G$ is defined by

$$\text{FSC}(k) = \frac{\sum_{|\vec{k}'|=k} F(\vec{k}')G^*(\vec{k}')}{\sqrt{\sum_{|\vec{k}'|=k} |F(\vec{k}')|^2} \sqrt{\sum_{|\vec{k}'|=k} |G(\vec{k}')|^2}}, \tag{30}$$

where $|\vec{k}'|$ is the magnitude of the spatial frequency vector $\vec{k}'$. FSC measures the similarity between the two maps. The closer the FSC is to one, the more similar the two normalized maps are. Since the Cryo-EM refinement is carried out in unsupervised fashion, the model quality is judged by examining the consistency of models across independent refinements. The gold standard FSC is calculated as following. At the beginning of refinement, the data is randomly split into two subsets with the same number of particles. Two sets of models are refined independently for each subset. The gold standard FSC refers to the FSC between these two independent reconstructions. The frequency where the gold-standard FSC curve passes through 0.143 is often denoted as the estimated resolution of the reconstructions [30]. Due to the noise outside the region where molecule resides, the FSC between two maps may underestimate the true resolution. In practice, we often use a mask to exclude contents outside the molecule and obtain a masked FSC between two masked maps. This result can also be used as a reference for the resolution estimation.

## 2.4    Model-Map FSC

If there exists a predetermined high resolution atomic model using X-ray crystallography, another way to validate the resolution of the Cryo-EM experimental map is to compare it with the atomic model. The resolution is determined by the correlation between atomic model map and Cryo-EM experimental map. The first step to calculate the model-map FSC is fitting the atomic model into the experimental density map. To avoid overfitting, we often employ the fitting method with minimal parameters such as rigid-body fitting. Then the model map is constructed from the fitted atomic model by sampling on the same grid as the experimental map. The model-map fit can then be evaluated by calculating the correlation between Fourier coefficients of model map and experimental map, namely, FSC, as it is defined in equation (30). This kind of FSC is referred as model-map FSC [31]. The point where the model-map FSC approaches a certain threshold can be referred to as the resolution of experimental map.

## 2.5 Parameter settings

To choose good parameters for our new method, we should determine the correct scale for our parameters. A theoretically optimal scale can be obtained by using the closed form solution of our new target function. For the lasso type problem, the closed form solution can be derived from its dual form [32]. The first step towards the dual form of our new target function is converting our new target function to a matrix form. With equation (26) in hand, we can write our 3D reconstruction problem in matrix form as

$$\min_{x} \frac{1}{2} \|y - DAx\|^2 \tag{31}$$
$$+ \alpha \sum_{j=1}^{L} \frac{|x_j|}{|x_j^i| + \epsilon} + \beta \sum_{j=1}^{L} \frac{\left\|\nabla x_j\right\|_2}{\left\|\nabla x_j^i\right\|_2 + \epsilon'} + \gamma \|x - x^{k-1}\|_2^2,$$

where $y$ is a vector representation of the 3D fourier transform data with $y(n) = \frac{\sum_{i=1}^{N} \sum_{j=1}^{L} \sum_{\phi} P(\phi|X_i, V_0) \kappa\left(\frac{\left\|n_j(\phi) - n\right\|}{h}\right) \mathrm{CTF}_{ij} X_{ij}}{\sqrt{N(n)}}$ for $n = [h, k, l] \in [0, L]$, $D$ is a diagonal matrix with diagonal element $D(n, n) = \sqrt{N(n)}$, and $Ax$ is the fourier transform of the 3D model $x$. We can derive the dual form of equation (31) by simplifying our restraint. According to [32], substituting the restraint in equation (31) by $\lambda \|Lx\|_1$, the dual of equation (31) with new restraint is of the form,

$$\min_{u} (A^T Dy - L^T u)^T (A^H D^2 A)^+ (A^T Dy - L^T u) \tag{32}$$

subject to $\|u\|_\infty \le \lambda$, $L^T u \in \mathrm{row}(DA)$, where $(A^T D^2 A)^+$ is the Moore-Penrose inverse for $A^H D^2 A$, $\|u\|_\infty = \max_{u_i \in u} |u_i|$, $\lambda$ is the parameter for $l_1$ restraint and $u$ is the dual variable of the 3D volume $x$. Given $u$, the closed form of the solution of $x$ can be written as,

$$x = (A^H D^2 A)^+ (A^H Dy - L^T u). \tag{33}$$

In equation (33), $A^H Dy$ is the inverse fourier transform of the data, which is the unregularized solution. $L^T u$ is the dual variable of restraint, which regularizes $A^H Dy$. To achieve sparsity in the solution $x$, $u$ needs to zero out certain components of $A^H Dy$. Since the dual variable $u$ is bounded by $\lambda$, the restraint parameter $\lambda$ should be of the same scale as the average of the magnitudes of $A^H Dy$. Though our restraint is of a more complex form than $\|Lx\|_1$, the dual of our restraint is in the space of a combination of two domains similar to $\|u\|_\infty \leq \lambda$. Detailed derivations about the dual space of the combination of two norm can be found in [33], [34]. Therefore, we can set the parameters of our restraint to be of the scale as the square root of the average of the squares of $A^H Dy$. We denote the square root of the average of the squares of $A^H Dy$ as $\overline{\|A^H Dy\|_2}$ in remaining sections.

The scale of implicit gradient descent restraint is easy to set since it is quadratic. Note that each quadratic data loss term in our target function is scaled by $N(n)$ in equation (28). Using the heuristic that the penalty term should match the loss term, we can set the restraint parameter $\gamma$ to be on the scale of the average of $N(n)$. We denote the average of $N(n)$ as $\overline{N(n)}$ in remaining sections.

Other important parameters to be set are $\epsilon$ and $\epsilon'$. They are in the denominators of our sparsity and smoothness restraint. If they are too small comparing with the density values of the 3D volume, the weights in our weighted norms will be very flexible and strongly depend on the magnitude of the value of each voxel in the 3D volume. Such kinds of restraints might not be able to effectively remove background noises and cause two independent refinements diverge. If they are too large comparing with the density values of the 3D volume, the restraints degrade to the original $l_1$ and TV norms and leads to more biased solutions. Optimal values of $\epsilon$ and $\epsilon'$ should assign large weights to background noises and small weights to true molecular densities. Hence, we can set the $\epsilon$ to the level of density values corresponding to molecular content in the 3D volume. $\epsilon'$ can be set close to $\epsilon$. This level can be easily obtained from the intermediate volumes generated by the refinements using traditional method. This is also similar to choose a threshold for creating a mask when computing masked FSC.

In conclusion, we should set the restraint parameters $\alpha$ and $\beta$ to be on the same scale as $\overline{\|A^H Dy\|_2}$. Since the corresponding restraints are inversely weighted by quantities with two other parameters $\epsilon$ and $\epsilon'$, we multiply the scale $\overline{\|A^H Dy\|_2}$ by $\epsilon$ or $\epsilon'$ to counter acting the effects of $\epsilon$ and $\epsilon'$. For zero elements, their restraint parameters are then normalized to be on the scale of $\overline{\|A^H Dy\|_2}$.

We performed a grid search on $\beta$-galactosidase to determine the possible ranges of parameters. We started by setting $\epsilon$ to 0.1, which is higher than the level of density values corresponding to molecular density. We considered setting $\epsilon'$ to be $\epsilon/3$ since the magnitude of gradient is often smaller than the density value of molecule. The initial guesses for $\alpha$ and $\gamma$ were $\alpha = 0.5\|\overline{A^H Dy}\|_2\epsilon$, $\gamma = 0.05\overline{N(n)}$. We scanned through $\beta \in [1, 2.2]\|\overline{A^H Dy}\|_2\epsilon'$ with a step size $0.2\|\overline{A^H Dy}\|_2\epsilon'$. The resolutions for every $\beta$ are shown in Figure 1. We can see that the best result can be found at $\beta = 1.6\|\overline{A^H Dy}\|_2\epsilon'$ or $\beta = 1.8\|\overline{A^H Dy}\|_2\epsilon'$.

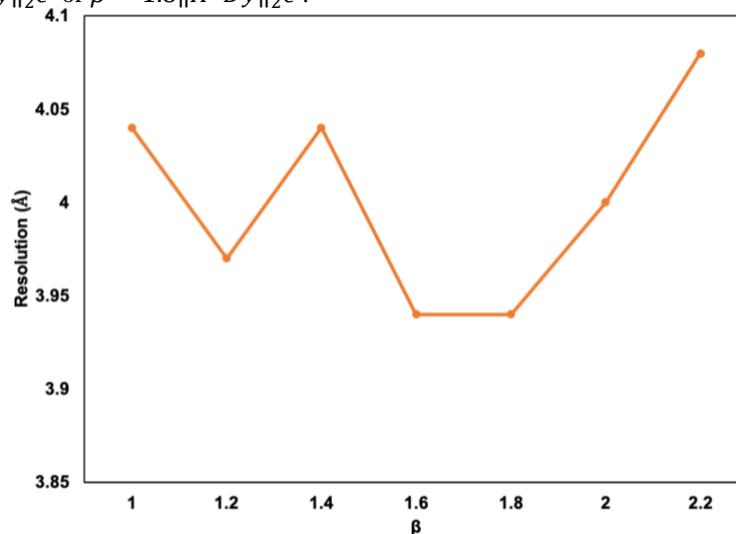

**Figure 1** Resolutions of results refined with different $\beta$ for $\beta$-galactosidase

For 80S ribosome, we used the same $\epsilon$, $\epsilon'$ and $\gamma$ as $\beta$-galactosidase. We scanned through $\beta \in [1.2, 2.4]$ with a step size $0.4\|\overline{A^H Dy}\|_2\epsilon'$ for $\alpha = 0.5\|\overline{A^H Dy}\|_2\epsilon$ and $\alpha = 0.4\|\overline{A^H Dy}\|_2\epsilon$. The results are shown in Figure 2. The best results were obtained at $\alpha = 0.4\|\overline{A^H Dy}\|_2\epsilon$ and $\beta = 1.6\|\overline{A^H Dy}\|_2\epsilon'$ or $2\|\overline{A^H Dy}\|_2\epsilon'$. For influenza HA trimer, we set $\epsilon$ to 0.015 and $\epsilon'$ to $\frac{\epsilon}{3}$. We increased $\gamma$ to $0.2\overline{N(n)}$. The $\alpha$ is fixed at $0.4\|\overline{A^H Dy}\|_2\epsilon$ during search. The resolutions of results for different $\beta$ are shown in Figure 3. The best result was obtained at $\beta = 2.6\|\overline{A^H Dy}\|_2\epsilon'$. We also reran the refinement after decreasing $\gamma$ to $0.1\overline{N(n)}$ while fixing other parameters we found in the best result. The resolution of this attempt decreased to 3.99Å. Hence, we stopped our search for HA trimer. Similar searches were applied to other cases and the best parameter are reported.

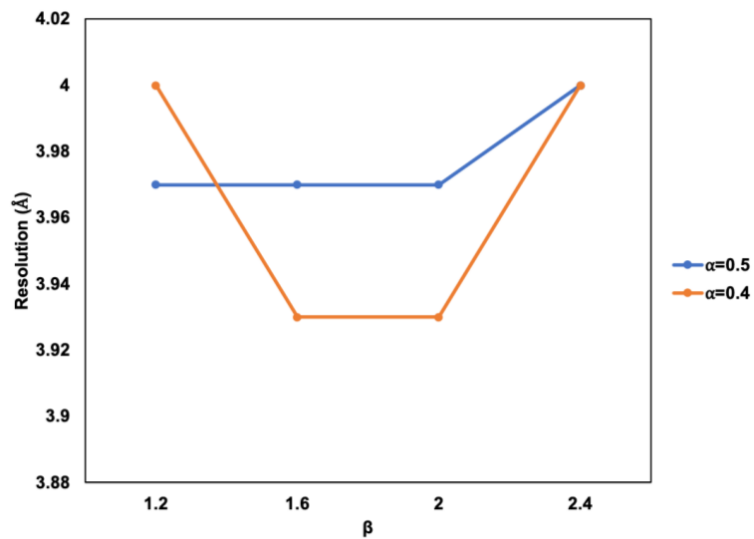

**Figure 2** Resolutions of results refined with different $\beta$ and $\alpha$ for 80S ribosome.

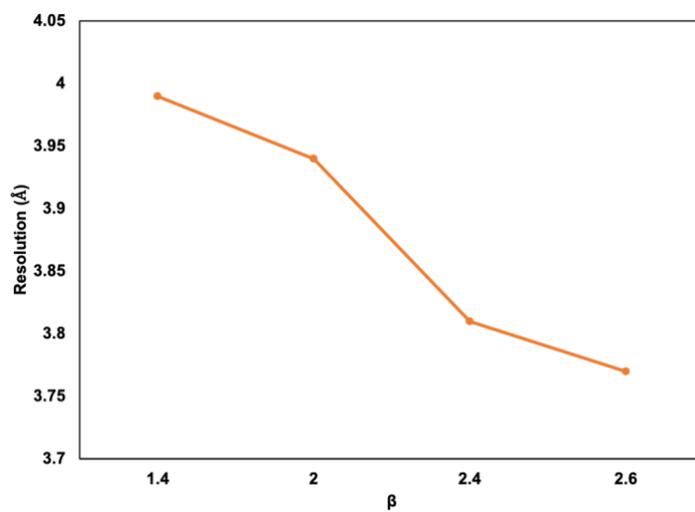

**Figure 3** Resolutions of results refined with different $\beta$ for influenza HA trimer.

## 3    Results

We tested our method by performing 3D refinement on real datasets and comparing the refinement results with the models obtained using traditional method on the same datasets. The 3D refinement process consists of iterations alternating between expectation and maximization. Both methods used the same settings such as adaptive sampling rate and oversample order in the expectation steps [13]. We also used the same convergence criteria for two methods during comparison, i.e., no resolution improvement and orientation and translation changes for at least two iterations [13]. The resolutions of different density maps refined by different methods are summarized in Table 1.

We used $\beta$-galactosidase as a test case. This data set has been extensively used in previous research [14], [35]. Since there was no ready-to-use particle stack for model building, our test began with extracting particles from micrographs using the coordinates manually picked by Richard Henderson [35]. After a round of 2D classification, the particles belong to the major classes were selected for model building. An initial model was generated *ab initio* from the 3D classification procedure. The initial model was then low pass filtered to 50 Å and averaged using D2 symmetry. We performed 3D refinements using different methods and the same initial model while enforcing D2 symmetry during the refinements. The best parameters of our method are $\epsilon = 0.1, \epsilon' = \frac{\epsilon}{3}, \alpha = 0.5\overline{\|A^H Dy\|_2}\epsilon, \ \beta = 1.8\overline{\|A^H Dy\|_2}\epsilon'$ and $\gamma = 0.05\overline{N(n)}$. In the post process step, we created a mask with the final reconstruction using all particles in the 3D refinement procedure. Using *relion_postprocess* routine [16], we obtained post processed maps from independent maps by correcting the MTF of the detector and sharpening with automatically estimated B-factors. To calculate the model map FSCs and model map correlations, we fitted the atomic coordinates of an *E. coli* $\beta$-galactosidase structure 3I3E [36] into the post processed density maps reconstructed by different methods using rigid boy fit in *Chimera* [37]. The results obtained from different methods are presented in Figure 4 and Figure 5. The FSC curves for the density maps refined by our new method are plotted with solid lines, and the FSC curves for the density maps refined by the traditional method are plotted with dash lines. The FSC curves between unmasked maps are colored in blue, and the FSC curves between masked maps are colored in red. The light blue line represents the FSC value equals 0.143. As it is shown in the Figure 4, the solid lines are above the dash lines in almost all resolution range. This suggests that the density maps obtained from our new method are of higher quality comparing with the density maps obtained by traditional method. The improvement of the density maps obtained from our new method is further validated by the model-map FSC curves calculated by *Phenix.Mtriage* [38]. The density maps reconstructed by our new method have higher correlations with respect to the rigid body fitted model in medium to high resolution shells.

We also tested our method on the 80s ribosome which is collected by [39]. Since no particle stack was deposited for this protein, our test was started from scratch. We extracted particles from the micrographs using the semi-automated selection process in Relion [40]. The particles were pruned by a round of 2D classification where only the particles classified to major classes were kept. We then constructed an *ab initio* model for 3D refinement through 3D classification as before. 3D refinements continued from the 70 Å low pass filtered initial model. The best parameters of our method are $\epsilon = 0.1, \epsilon' = \frac{\epsilon}{3}, \alpha = 0.4\overline{\|A^H Dy\|_2}\epsilon, \ \beta = 1.6\overline{\|A^H Dy\|_2}\epsilon'$ and $\gamma = 0.05\overline{N(n)}$. A post process which is similar to the preceding case was applied on the final density maps. To assess the Cryo-EM maps determined using different methods, we fitted 80S crystal structure, 3U5B [41], using a simple rigid body fit into postprocessed maps to obtain high resolution reference atomic structures. The 40S and 60S subunits were fitted separately. The model-map FSCs were calculated between our maps and corresponding reference structures. Using the same color scheme and line type as in the $\beta$-galactosidase figures, the gold standard FSC curves for refinements using different methods are reported in Figure 6. The FSC curves between density maps from our new method are above the FSC curves between density maps from the traditional method. For the model-map FSC curves shown in Figure 7, the FSC between density map refined by our method and the rigid body fitted model overlaps with the FSC between density map obtained by traditional method and the rigid body fitted model in low to medium resolution. However, the density map refined by our method has higher correlation with the model map in high resolution regions, which suggests that the density map obtained by our method achieves higher resolution comparing with the density map refined by traditional method.

We tested our method on the influenza hemagglutinin (HA) trimer. The data was obtained from EMPIAR deposition with accession number 10097 [42]. We generated an initial model *ab initio* using the 3D classification. The initial model was further averaged according to C3 symmetry. The 3D refinements were performed by using a 40 Å low-passed filtered initial model and enforcing the C3 symmetry. The best parameters of our method are $\epsilon = 0.015, \epsilon' = \frac{\epsilon}{3}, \alpha = 0.4\overline{\|A^H Dy\|_2}\epsilon, \ \beta = 2.4\overline{\|A^H Dy\|_2}\epsilon'$ and $\gamma = 0.2\overline{N(n)}$. To compare the results obtained from two different refinement methods, we used a high resolution atomic structure of HA trimer (PDB:3WHE) which was determined by Xray crystallography as a reference [43]. The atomic model was rigidly fitted into Cryo-EM maps using *Chimera* [37]. The model-map FSC curves were reported. We also compared the post processed maps derived from the results of different refinement methods. In the post process step, the final reconstruction using all particles in the 3D refinement procedure was used to generate a mask. The final density maps were created from the independent maps and corrected for the modulation transfer function (MTF) of the detector using *relion_postprocess* routine. They then sharpened by applying a negative B-factor that was automatically estimated. The gold standard FSC curves between density maps refined by different methods are plotted in Figure 8 using the same color scheme and line type as before. The unmasked gold standard FSC of our new method is greatly improved over the unmask gold standard FSC of traditional method. The masked gold standard FSC of our new method is also improved in most of resolution shells. The density map refined by our new method has higher resolution according to the 0.143 criterion. This improvement is further confirmed by the model-map FSCs in Figure 9. It is easy to see that the post processed map of our new method has much higher correlation w.r.t the atomic model in medium to high resolution shells.

Another test case is the structure of the protein-conducting ERAD channel Hrd1 in complex with Hrd3 [44]. The Cryo-EM data was downloaded from the EMPIAR with accession number 10099. Due to the heterogeneity of the dataset, the 3D classification was used to classify the particles, and generate the corresponding initial models of different complexes for 3D refinements. The particles which were classified as Hrd1/Hrd3 dimer were selected, and then subject to 3D refinements. We performed 3D refinements using different methods and the same 20 Å low-passed filtered initial model. The best parameters of our method are $\epsilon = 0.01, \epsilon' = \frac{\epsilon}{2}, \alpha = 0.4\overline{\|A^H Dy\|_2}\epsilon$, $\beta = 1.2\overline{\|A^H Dy\|_2}\epsilon'$ and $\gamma = 0.1\overline{N(n)}$. The final density maps were created using same approaches as before. The final results were compared w.r.t the atomic models of Hrd1 dimer (5V6P) and Hrd3 monomer (5V7V) by calculating model-map FSC [44]. The Hrd1 dimer and Hrd3 monomer were fitted into density map separately. The gold standard FSC curves between the density maps refined by different methods are shown in Figure 10. It is easy to see that the FSC curves between density maps refined by our new methods is higher than the FSC curves between density maps refined by traditional method in most regions, which suggests that the refinement results of our new method should have higher resolution comparing with the refinement results of traditional method. The model-map FSC curves in Figure 11 corroborates our conclusion since the density map obtained by our method again has higher correlation w.r.t the rigid-body fitted model map in most resolution shells. Therefore, the density map obtained by our new method achieved higher resolution comparing to the density map obtained by traditional method.

The final test case is the structure of the TMEM16A calcium-activated chloride channel [45]. The Cryo-EM data was downloaded from the EMPIAR with accession number 10123. We generated an initial model *ab initio* using the 3D classification. The initial model was further averaged according to C2 symmetry. The 3D refinements were performed by using a 40 Å low-passed filtered initial model and enforcing the C2 symmetry. The best parameters of our method are $\epsilon = 0.01, \epsilon' = \epsilon, \alpha = 0.5\|\overline{A^H Dy}\|_2\epsilon, \ \beta = 2\|\overline{A^H Dy}\|_2\epsilon'$ and $\gamma = 0.1\overline{N(n)}$. The final density maps were created using same approaches as before. The final results were compared w.r.t the atomic model 6BGI by calculating model-map FSC [45]. The gold standard FSC curves between the density maps refined by different methods are shown in Figure 12. It is easy to see that the FSC curves between density maps refined by our new methods is higher than the FSC curves between density maps refined by traditional method in most regions, which suggests that the refinement results of our new method should have higher resolution comparing with the refinement results of traditional method. Even though the atomic model was obtained by fitting against the density map refined by traditional method, we can still observe that the model-map FSC curves of our method is above the same kinds of model-map FSC curves of traditional method in all resolution shells in Figure 13. This corroborates our conclusion that the density map obtained by our new method achieved higher resolution comparing to the density map obtained by traditional method.

We visually inspected different final density maps reconstructed by different refinement methods, and compared them with the corresponding atomic models. We here present some differences between density maps refined by different methods along with the structures. The first example is for HA trimer and located within the residue 451 to 455 in chain C. The density map refined by our new method is plotted in Figure 14 (a), and colored in green. The density map refined by traditional method is plotted in Figure 14 (b), and colored in red. Both density maps are contoured at the same level. It is easy to see that the density map refined by traditional method has no density at the sidechain of residue 452 and incomplete density at the sidechain of residue 453, while the density map refined by our new method presents more complete density at these residues. Besides, the sidechain of residue 454 stays outside the density map refined by traditional method, while it is enclosed by the density map refined our new method. These differences suggest that the density map refined our new method has higher map model correlation than the density map refined by traditional method. Another example is found in residues from 386 to 391 in the chain A of Hrd1/Hrd3 dimer. The density maps refined by different methods are plotted along with the corresponding structure and colored by the same color scheme in Figure 15. The density map refined by our new method shows density for the sidechain of residue 391, while the density map refined by traditional method has no density above this level in the same region. Besides the density map refined by our new method covers the sidechains of residue 389 and 390, while some atoms in those sidechains stay outside the density map refined by traditional method. These differences serve as complementary evidences to indicate that our new method can improve the model map correlation of the final density map. For $\beta$-galactosidase, residues between 796 and 801 in chain A are taken as an example. We can also observe some improvements in the density map refined by our new method. As it is shown in Figure 16, the density map refined by our new method has more densities for the sidechain of residue 799 in contrast to the density map refined by traditional method. The whole sidechain of residue 796 is surrounded in the density map refined by our new method, while there are few atoms stay outside of the density map refined by traditional method. The last difference is that the density map refined by our new method has more density near the sidechain of residue 800. The final example is taken from the residues between 105 and 120 in the chain J of 80S ribosome. As it is shown in Figure 17, the most noticeable improvements of density map refined by our new method in this region are located at residue 109 and residue 114. Contoured at the same level, our density map shows clear density for sidechains of these residues comparing with the density map refined by traditional method. We can also observe that our density map has more complete density for the sidechains of residue 120, 117 and 116. Finally, we presented the two post processed density maps of TMEM16A in Figure 18. The density map of TMEM16A refined by our new method is shown in Figure 18 (a), while the density map of TMEM16A refined by traditional method is shown in Figure 18 (b). Both maps are contoured at the same level. We can see that the density map refined by our new method exhibits more secondary structure features in the middle of this density map comparing the density map refined by traditional method. Besides, there are also more densities at the top and bottom of the density map refined by our new method.

In all results presented in this paper, we can see that the improvement of our new method in gold standard unmasked FSC is most noticeable. This phenomenon suggests that our method has superior denoising effect to the 3d volume, thus producing less noisy reference model during refinement. The cleaner model in turn leads to more accurate orientation and translation parameter estimation for each image in the expectation step. These reciprocal improvements of our method result in better refinement results eventually. Moreover, though our new method introduced more parameters, the parameter settings of our new method which generates better result than traditional method can be easily found in a relatively small range. First of all, $\epsilon$ can be set according to the density level of molecular content in the 3D volume. This is of the same fashion as selecting the threshold for creating mask when computing masked FSC. In our test cases, $\alpha$s are in the range $[0.4,0.5]\overline{\|A^H Dy\|_2}\epsilon$, $\beta$s are in the range $[1.2,2.4]\overline{\|A^H Dy\|_2}\epsilon'$, and $\gamma$s are in the range $[0.05,0.2]\overline{N(n)}$. $\epsilon'$s are in the range $\left[\frac{1}{3},1\right]\epsilon$. These ranges provide useful guidance for future applications of our new method. Given these ranges, we can use the brute force approach – grid search to obtain the optimal parameter setting.

## 4 Conclusion

In this paper, we proposed a new type of 3D refinement method for the Cryo-EM single particle analysis. Our analysis reveals that our new method promotes different kinds of smoothness comparing with the traditional method. Unlike the traditional method [13] which enforces translation invariant rotationally symmetric smoothness to the 3D model, our new method enforces spatially varying smoothness to the 3D model. Since most structures does not exhibit rotational symmetry, the traditional method might result in large bias in the final model. In contrast, our method can adapt to different smoothness in different regions in structures, thus greatly reducing model biases and improving the final results. Another approach to promote smoothness we explored in this paper is by formulating the backprojection as a local kernel regression problem. This new formulation enables us to embed the 3D model in a RKHS with specific smoothness. We also introduced a new prior, sparsity, into the 3D refinement process. By setting relatively small values in the 3D model to zero, the sparsity restraint can suppress the strength of signal which doesn't belong to molecules in the 3D model and leads to better refinement result. We tested our new method on real datasets and compared the refinement results with the results obtained by traditional method. Using the criteria like gold standard FSC and model map FSC, we have shown that the results obtained by our new method has greatly improved the resolution of electron density map and the correlation between the atomic model and electron density map. We expect our method to be deployed in Cryo-EM structure determination process and help experimenters to obtain higher resolution maps and structures.

**Table 1** Resolutions of different density maps of proteins refined by different methods. Gold Standard Traditional represents the resolution of the density map refined by traditional method. Gold Standard Our represents the resolution of the density map refined by traditional method. Both resolutions are measured at FSC = 0.143 using the gold standard FSC and phase randomisation method [16]. Model Map Traditional refers to the resolution of density map refined by traditional method. Model Map Our refers to the resolution of density map refined by our method. Both resolutions are obtained at FSC = 0.143 using masked model-map FSC.

| Proteins | Gold Standard Traditional | Gold Standard Our | Model Map Traditional | Model Map Our |
|---|---|---|---|---|
| $\beta$-galactosidase | 4.16 | 3.97 | 4.05 | 3.91 |
| 80S ribosome | 4.08 | 3.93 | 4.04 | 3.89 |
| hemagglutinin | 4.19 | 3.77 | 4.06 | 3.72 |

| | | | |
|---|---|---|---|
| Hrd1/Hrd3 | 4.80 | 3.55 | 4.70 | 4.25 |
| TMEM14A | 4.01 | 2.93 | 3.87 | 3.14 |

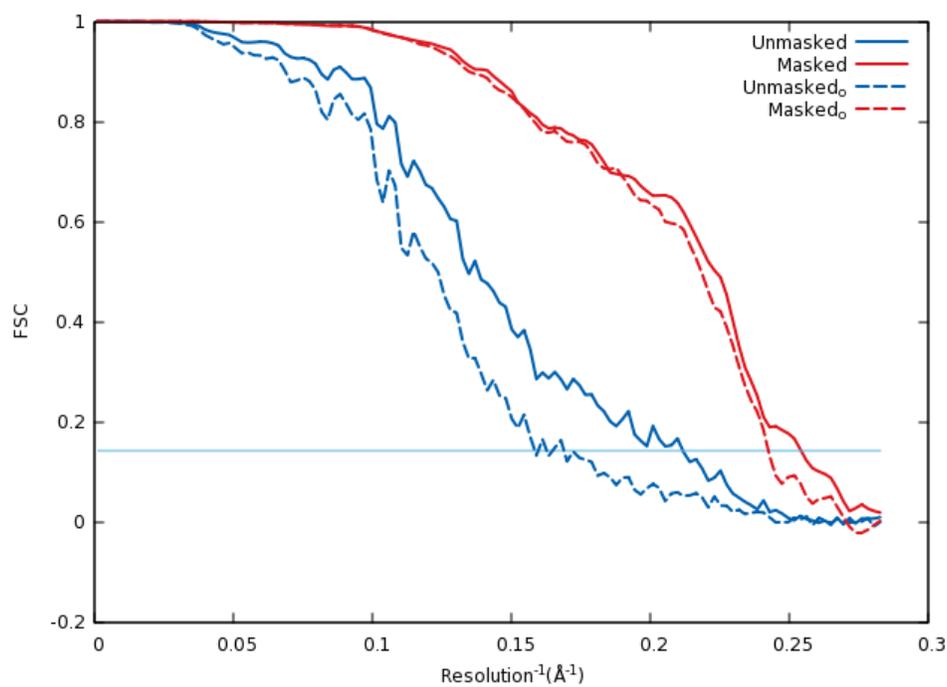

**Figure 4** Gold standard unmasked FSC curves and masked FSC curves of the $\beta$-galactosidase calculated from two independent reconstructions for different refinement methods. Unmasked represents the unmasked FSC curve of results from our method, and masked represents the masked FSC curve of results from our method. $Unmasked_o$ represents the unmasked FSC curve of results from traditional method, and $masked_o$ represents the masked FSC curve of results from traditional method.

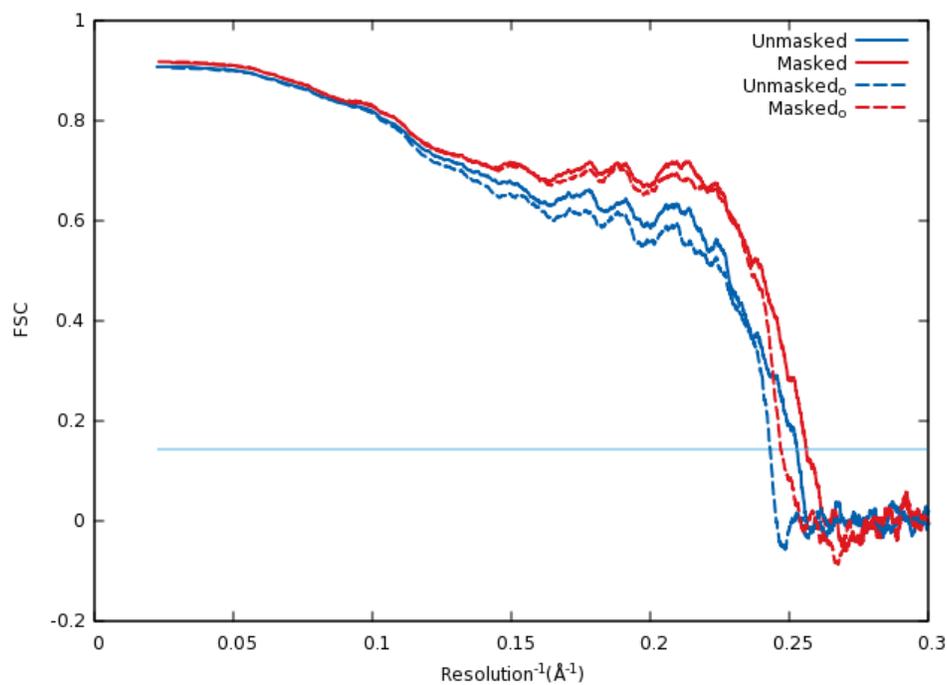

**Figure 5** Model-map FSC curves between the post processed density maps of the $\beta$-galactosidase obtained using different methods and the corresponding rigid-body fitted atomic models 3I3E. Unmasked represents the unmasked FSC curve of results from our method, and masked represents the masked FSC curve of results from our method. $Unmasked_o$ represents the unmasked FSC curve of results from traditional method, and $masked_o$ represents the masked FSC curve of results from traditional method.

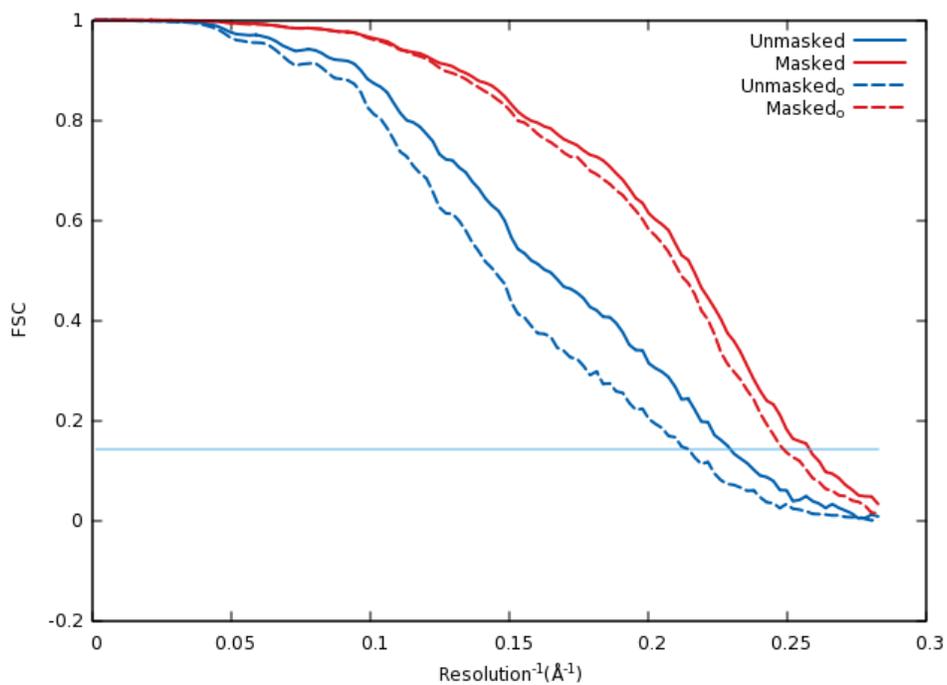

**Figure 6** Gold standard unmasked and masked FSC curves of 80S ribosome calculated between two independent reconstructions for different methods. Unmasked represents the unmasked FSC curve of results from our method, and masked represents the masked FSC curve of results from our method. $Unmasked_o$ represents the unmasked FSC curve of results from traditional method, and $masked_o$ represents the masked FSC curve of results from traditional method.

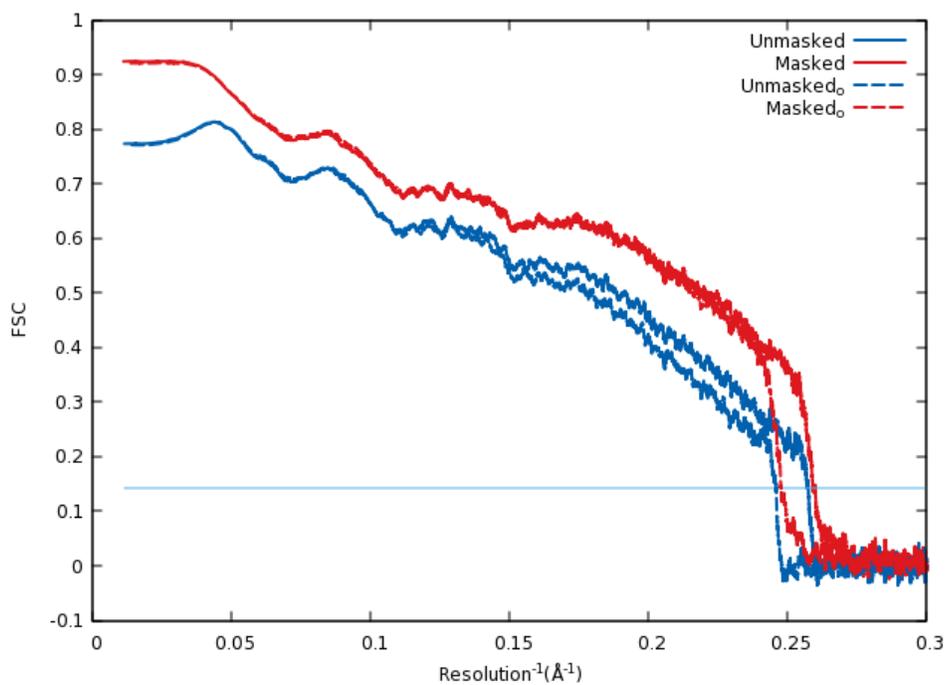

**Figure 7** FSC curves between the post processed maps refined using different methods and the corresponding rigid-body fitted atomic models 3U5B, for which the 40S and 60S subunits are fitted separately. Unmasked represents the unmasked FSC curve of results from our method, and masked represents the masked FSC curve of results from our method. $Unmasked_o$ represents the unmasked FSC curve of results from traditional method, and $masked_o$ represents the masked FSC curve of results from traditional method.

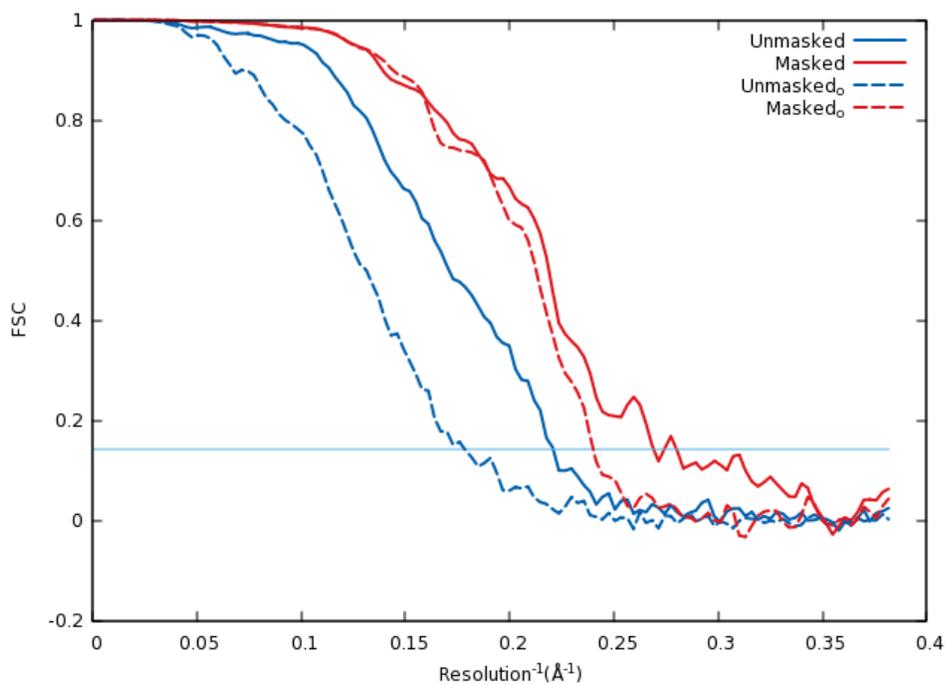

**Figure 8** Gold standard unmasked and masked FSC curves of the influenza hemagglutinin trimer between two independent refinements for different methods. Unmasked represents the unmasked FSC curve of results from our method, and masked represents the masked FSC curve of results from our method. $Unmasked_o$ represents the unmasked FSC curve of results from traditional method, and $masked_o$ represents the masked FSC curve of results from traditional method.

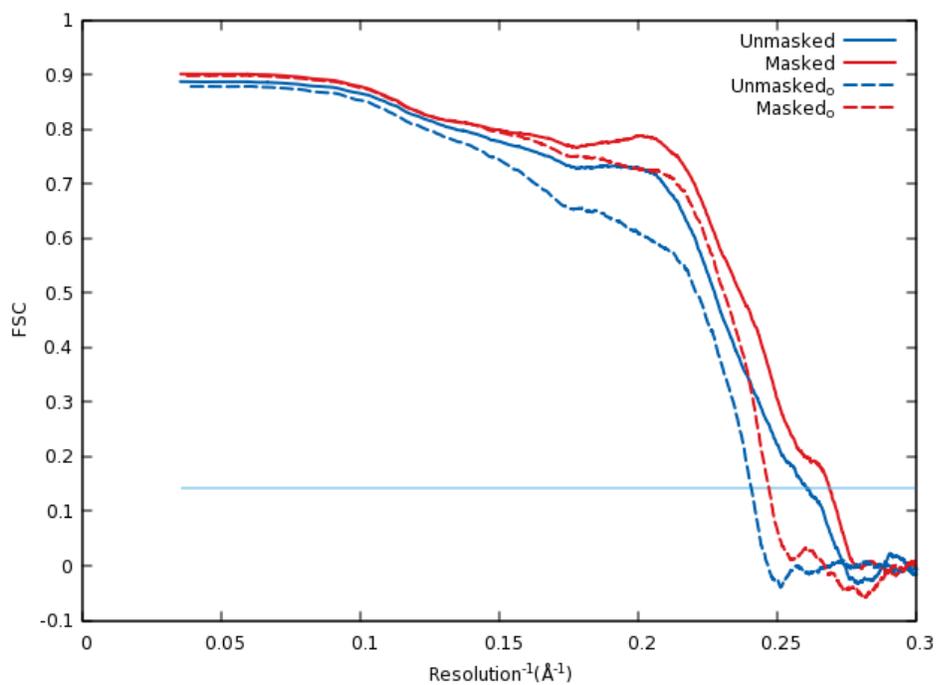

**Figure 9** FSC curves between the post processed maps refined using different methods and the corresponding rigid-body fitted atomic models 3WHE. Unmasked represents the unmasked FSC curve of results from our method, and masked represents the masked FSC curve of results from our method. $Unmasked_o$ represents the unmasked FSC curve of results from traditional method, and $masked_o$ represents the masked FSC curve of results from traditional method.

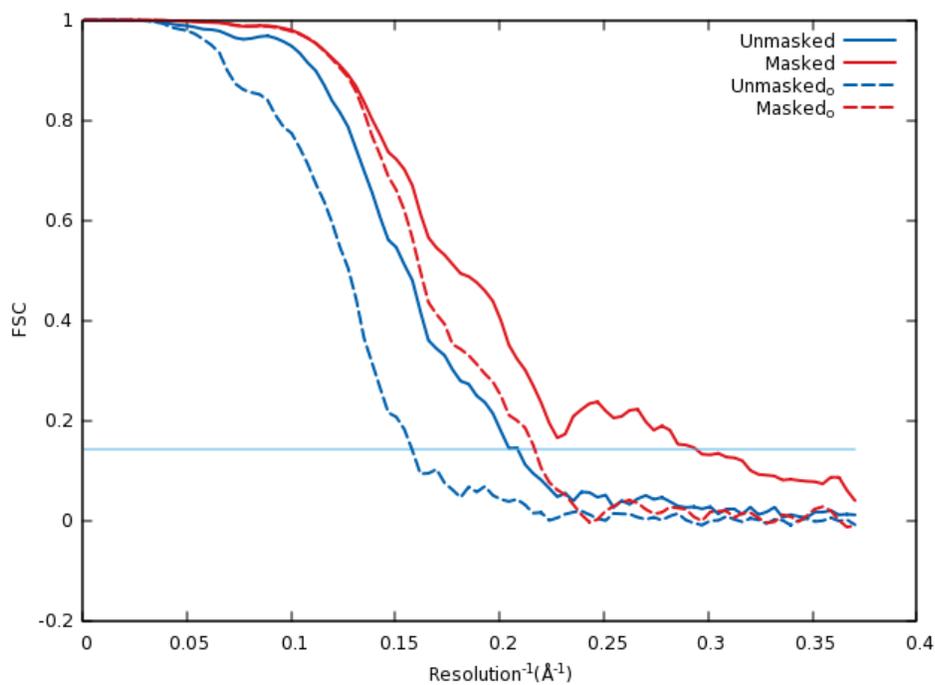

**Figure 10** Gold standard unmasked and masked FSC curves of the Hrd1/Hrd3 complex between two independent refinements for different methods. Curves with different colors show the results of different methods. Unmasked represents the unmasked FSC curve of results from our method, and masked represents the masked FSC curve of results from our method. $Unmasked_o$ represents the unmasked FSC curve of results from traditional method, and $masked_o$ represents the masked FSC curve of results from traditional method.

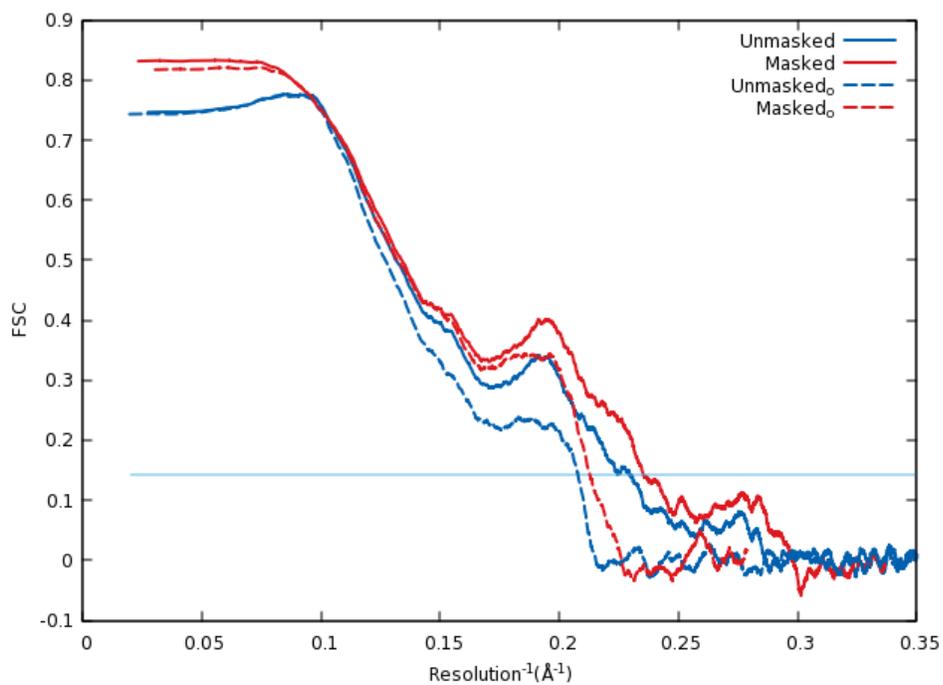

**Figure 11** FSC curves between the post processed maps refined using different methods and the corresponding rigid-body fitted atomic models, for which the Hrd1 dimer and Hrd3 subunits are fitted separately. Unmasked represents the unmasked FSC curve of results from our method, and masked represents the masked FSC curve of results from our method. $Unmasked_o$ represents the unmasked FSC curve of results from traditional method, and $masked_o$ represents the masked FSC curve of results from traditional method.

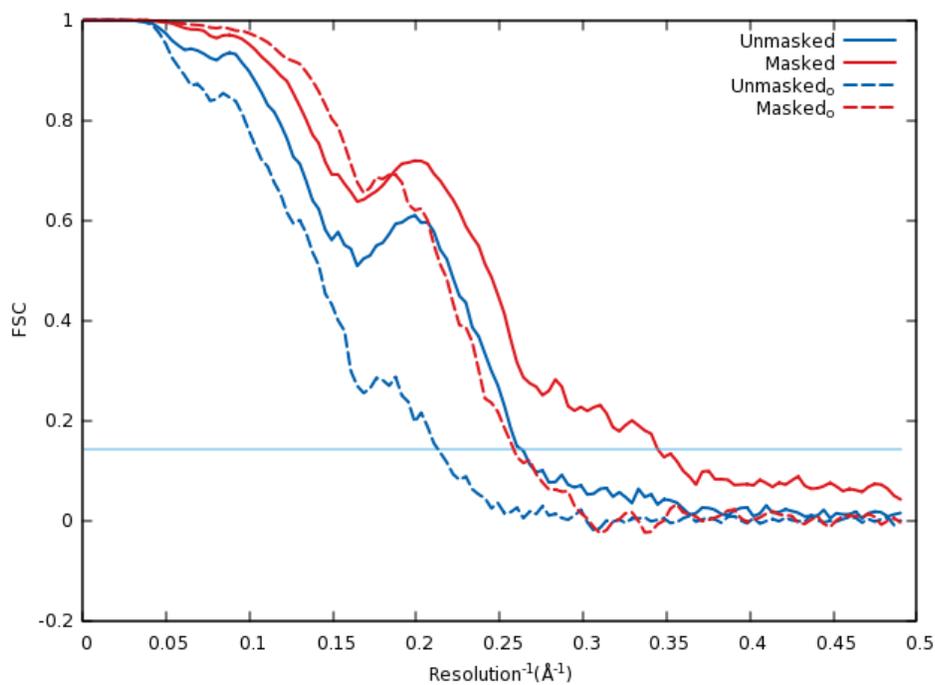

**Figure 12** Gold standard unmasked and masked FSC curves of the TMEM16A calcium-activated chloride channel between two independent refinements for different methods. Curves with different colors show the results of different methods. Unmasked represents the unmasked FSC curve of results from our method, and masked represents the masked FSC curve of results from our method. $Unmasked_o$ represents the unmasked FSC curve of results from traditional method, and $masked_o$ represents the masked FSC curve of results from traditional method.

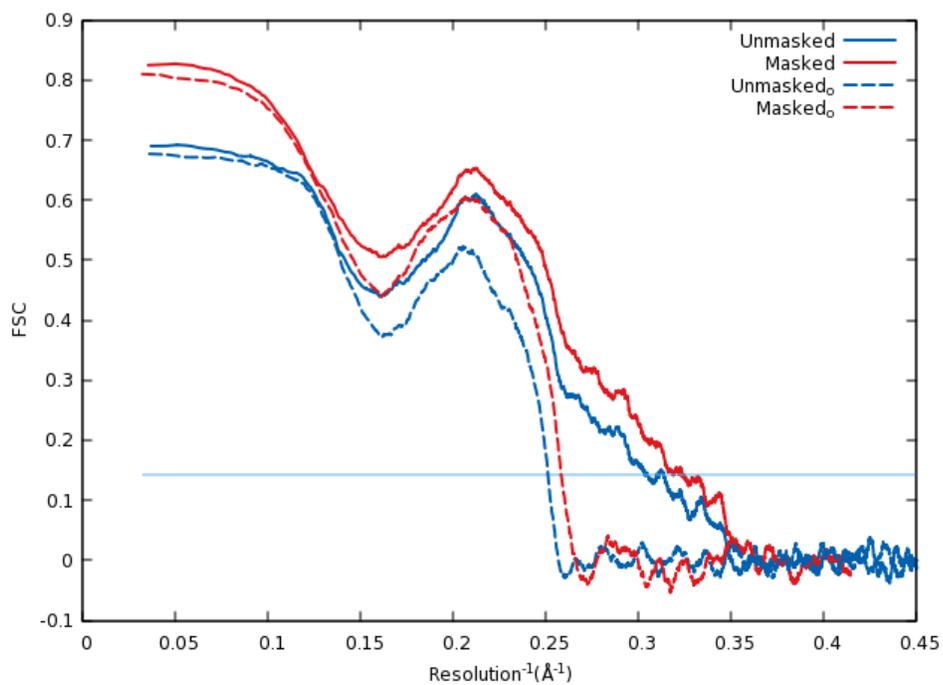

**Figure 13** FSC curves between the post processed maps refined using different methods and the corresponding rigid-body fitted atomic models. Unmasked represents the unmasked FSC curve of results from our method, and masked represents the masked FSC curve of results from our method. $Unmasked_o$ represents the unmasked FSC curve of results from traditional method, and $masked_o$ represents the masked FSC curve of results from traditional method.

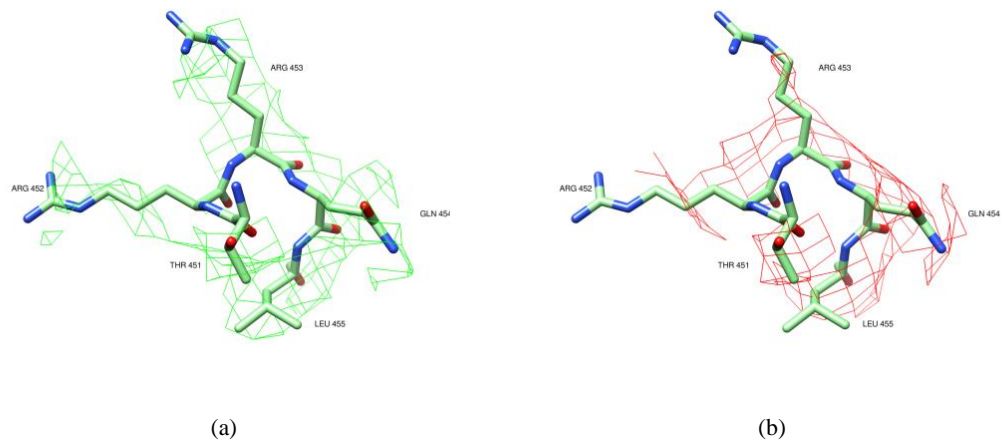

(a)                                    (b)

**Figure 14** (a) Density map refined by new method for residues from 451 to 455 in the chain C of HA trimer. (b) Density map refined by traditional method for residues from 451 to 455 in the chain C of HA trimer

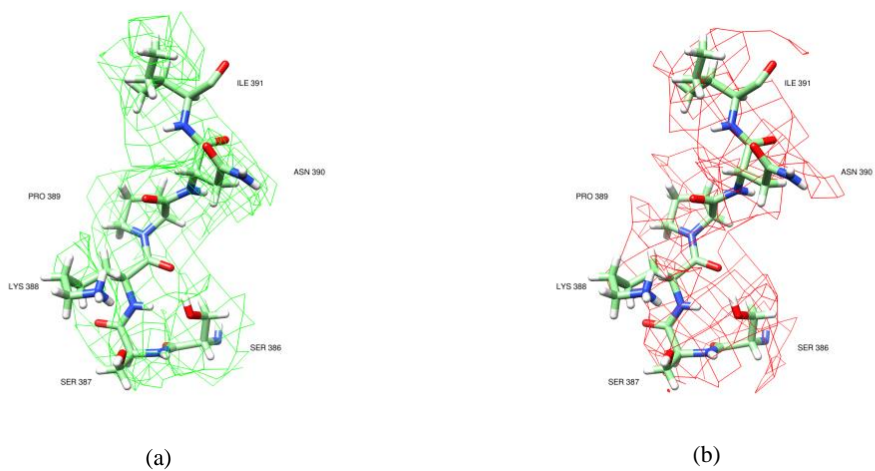

(a)                                    (b)

**Figure 15** (a) Density map refined by new method for residues from 386 to 391 in the chain A of Hrd1/Hrd3 dimer. (b) Density map refined by traditional method for residues from 386 to 391 in the chain A of Hrd1/Hrd3

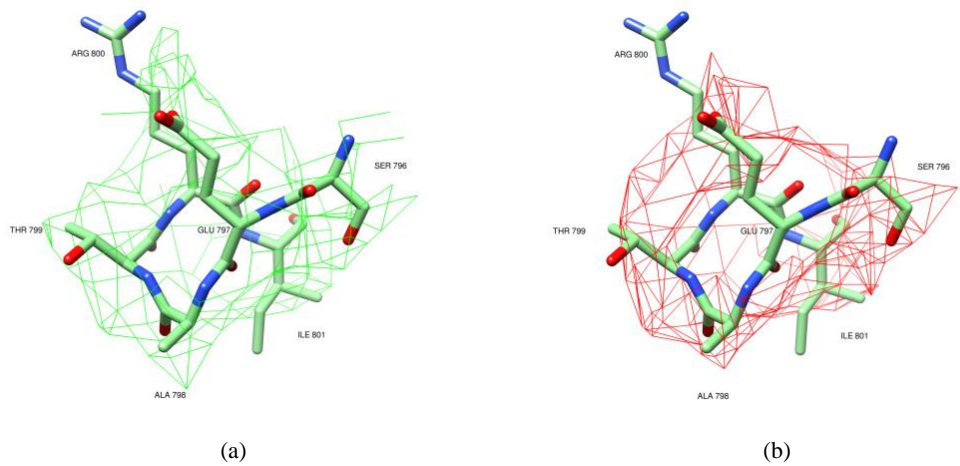

(a)                                    (b)

**Figure 16** (a) Density map refined by new method for residues from 796 to 801 in the chain A of β-galactosidase. (b) Density map refined by traditional method for residues from 796 to 801 in the chain A of β-galactosidase

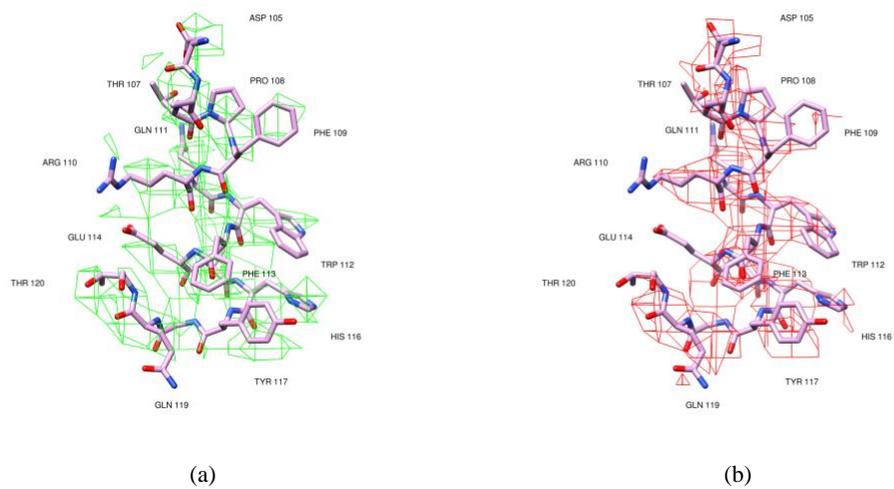

(a)                                    (b)

**Figure 17** (a) Density map refined by new method for residues from 105 to 120 in the chain J of 80S ribosome. (b) Density map refined by traditional method for residues from 105 to 120 in the chain J of 80S ribosome

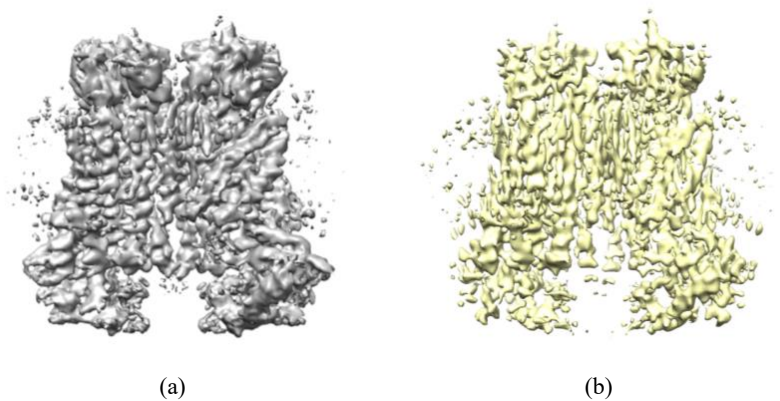

(a)                                        (b)

**Figure 18** (a) The postprocess density map for the TMEM16A refined by our method. (b) The postprocess density map for the TMEM16A refined by traditional method.